\documentclass[a4paper,12pt]{article}

\usepackage[dvips]{graphicx}
\usepackage{psfrag}
\usepackage{caption2}
\usepackage{amsmath}
\usepackage{amssymb}
\usepackage{amsfonts}

\title{Precession of Pericenter: A More Accurate Approach}
\author{Andreas Stergiou\footnote{e-mail: astergio@science.uva.nl}}
\date{}

\begin{document}

\maketitle

\begin{abstract}
In this paper we study the orbits of massive bodies moving in the
spacetime generated by a spherically symmetric and non-rotating distribution
of mass. More specifically, our treatment discusses the more accurate
calculation of the precession of pericenter due to general-relativistic
effects. Our result is accurate up to terms of second order, while
the precession met in the bibliography is accurate only up to first-order
terms.
\end{abstract}

\section{The Schwarzschild spacetime}
The first solution of Einstein's field equations was published by Karl Schwarz\-schild in 1916. It is a solution that informs us about the spacetime generated in the exterior of a spherically symmetric and non-rotating distribution of mass. As is well known, that distribution of mass provides the region that surrounds it with a static, spherically symmetric spacetime. That kind of spacetime is mathematically denoted by the line element (cf. \cite{Schutz})
\begin{eqnarray}\nonumber
ds^2 &=& g_{tt}(r)dt^2 + g_{rr}(r)dr^2 + r^2d\theta^2 +r^2 \sin^2\theta\, d\phi^2 \\
&=& -e^{2\Phi}dt^2+e^{2\Lambda}dr^2+r^2d\theta^2 +r^2 \sin^2\theta\, d\phi^2\nonumber
\end{eqnarray}
in spherical polar coordinates $(t,r,\theta,\phi)$. In the second step above we introduced the functions $\Phi\equiv\Phi(r)$ and $\Lambda\equiv\Lambda(r)$ in place of the two unknowns $g_{tt}(r)$ and $g_{rr}(r)$ respectively. That replacement was possible since $g_{tt}<0$ and $g_{rr}>0$ anywhere in spacetime.\footnote{Bear in mind, however, that this allegation breaks down in the case of black holes, where we should reconsider our system of coordinates.} Of course, we have to impose a couple of boundary conditions to the aforementioned line element, i.e.
$$\lim_{r\rightarrow\infty}\Phi(r)=\lim_{r\rightarrow\infty}\Lambda(r)=0$$
for we demand, as is physically reasonable, that spacetime be flat far away from the distribution of mass.

The previous results give us the ability to calculate the components $G^{\mu\nu}$ ($\mu, \nu = 0, 1, 2, 3$) of Einstein's tensor. The missing step, now, in order to calculate the unknown functions $\Phi(r)$ and $\Lambda(r)$, is to find out the components $T^{\mu\nu}$ of the stress-energy tensor, plug them into Einstein's field equations,
$$G^{\mu\nu}=8\pi T^{\mu\nu}$$
(note that here and hereafter we use geometrized units unless otherwise mentioned) and solve the resulting differential equations for $\Phi(r)$ and $\Lambda(r)$. The aforementioned tedious calculations lead us to the expressions\footnote{For further details see \cite{Schutz}}
$$e^{2\Phi} = e^{-2\Lambda}=1-\frac{2M}{r}$$
where $M$ is the total mass of the distribution of mass. Therefore, we are finally in position to write down the line element in its final form:
$$ds^2=-\left(1-\frac{2M}{r} \right)dt^2 + \left(1-\frac{2M}{r}
\right)^{-1} dr^2 + r^2\,d\theta^2 + r^2 \sin^2\theta\,d\phi^2$$

\section{Precession of Pericenter}
In order to calculate the precession of the pericenter of the elliptic orbit of a massive body revolving an attractive center, we begin from the equation of motion
\begin{equation}\label{part}
\left(\frac{dr}{d\tau}\right)^2=\tilde{E}^2-\left(1-
\frac{2M}{r}\right)\left(1+\frac{\tilde{L}^2}{r^2}\right)
\end{equation}
where we suppose that the elliptic orbit is described by the radial distance, $r$, and the azimuthal angle, $\phi$, $\tau$ is the proper time, $\tilde{E}$ and $\tilde{L}$ are the energy and momentum per unit mass respectively and $M$ is the total mass of the attractive center. The shape of the effective potential $\tilde{V}^2(r)=\left(1-\frac{2M}{r}\right)\left(1+\frac{\tilde{L}^2}{r^2}\right)$ for a typical value of the parameter $\tilde{L}^2$ is shown in Fig. 1.
\begin{figure}[ht]
\captionstyle{center}
\centering
\psfrag{V^2}[][]{$\tilde{V}^2$}
\psfrag{E^2=1}[][]{$\tilde{E}^2=1$}
\psfrag{0}[][]{$0$}
\psfrag{2M}[][]{$2M$}
\psfrag{A}[][]{$A$}
\psfrag{B}[][]{$B$}
\psfrag{r}[][]{$r$}
\includegraphics{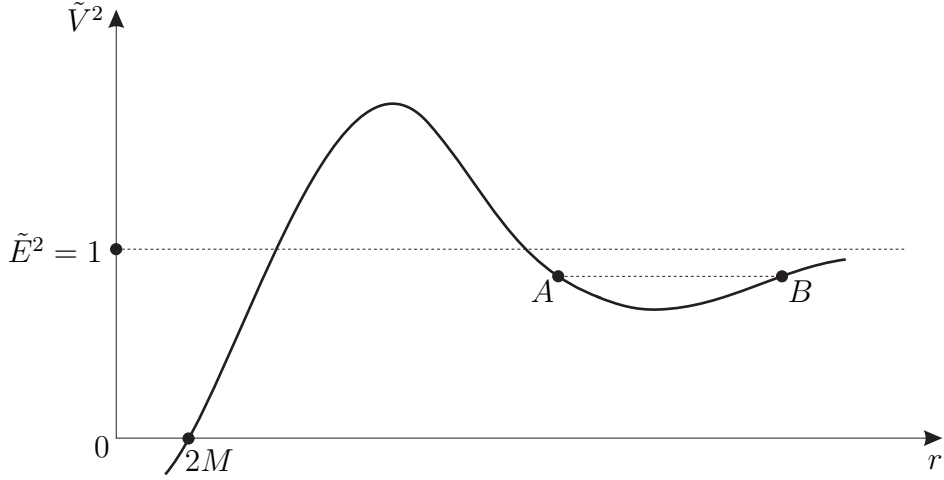}
\caption{Typical effective potential of a body with defined angular momentum in a Schwarzschild spacetime}
\end{figure}
Our study will be limited to energies $\tilde{E}^2 \approx 1$, for we want the body to be in a stable orbit around the attractive center. As an example we can see the points $A$ and $B$ in Fig. 1, among which the body can move in a nearly elliptic orbit.

 In order to write equation \eqref{part} in the more convenient form ${{dr}
\mathord{\left/ {\vphantom {{dr} {d\phi }}} \right.
\kern-\nulldelimiterspace} {d\phi }} = f(r)$ we use the fact that
$$\frac{{d\phi }}{{d\tau }} \equiv v^\phi   = \frac{{p_\phi  }}{m}
=g^{\phi \phi }
\frac{{p_\phi  }}{m}  =
\frac{1}{{r^2 }}\tilde{L}$$
Therefore, we get
$$\left( {\frac{{dr}}{{d\phi }}} \right)^2  = \frac{{\tilde{E}^2  -
\left( {1 - \frac{{2M}}{r}} \right)\left( {1 + \frac{{\tilde{L}^2
}}{{r^2 } }} \right)}}{\frac{\tilde{L}^2} {r^4} }$$
We now proceed with the transformation $u = \frac{{r_0 }}{r}$, where ${r_0 }$ is a constant with length dimensions. That transformation gives us a differential equation with no dimensions:
$$\frac{{du}}{{d\phi }} = \sqrt
{\frac{{\tilde{E}^2 - 1}}{{\tilde{L}^2 }}r_0^2  + \frac{{2Mr_0
}}{{\tilde{L}^2 }}u - u^2  + \frac{{2M}}{{r_0 }}u^3 }$$
It is clear that we can separate the variables in the last differential equation, thus taking
$$\frac{{du}}{{\sqrt {\frac{{\tilde{E}^2  -
1}}{{\tilde{L}^2 }}r_0^2  + \frac{{2Mr_0 }}{{\tilde{L}^2 }}u - u^2
+ \frac{{2M}}{{r_0 }}u^3 } }} = d\phi$$

Let, now, $a$, $b$ and $c$ be the roots of the polynomial
$$P(u) = \frac{{\tilde{E}^2  - 1}}{{\tilde{L}^2 }}r_0^2  +
 \frac{{2Mr_0}}{{\tilde{L}^2 }}u - u^2  + \frac{{2M}}{{r_0 }}u^3$$
Then, $P(u)$ can be written as $$P(u) = \alpha (u - a)(b - u)(c - u)$$
where $\alpha  = 2M/r_0$. Because of the equality of the coefficients of the same-order terms for the two forms of $P(u)$ we get the equations
\begin{equation}\label{coef-p-1}
 abc = \frac{{1 - \tilde{E}^2 }}{{2M\tilde{L}^2 }}r_0^3
\end{equation}
\begin{equation}\label{coef-p-2}
ab + c(a + b) = \frac{{r_0^2 }}{{\tilde{L}^2 }}
\end{equation}
and
\begin{equation}\label{coef-p-3}
a + b + c = \frac{{r_0 }}{{2M}}
\end{equation}

We suppose that within the range of values of energy for which we study our problem, $P(u)$ has the three real roots, $a$, $b$ and $c$ for which \mbox{$a < b \ll c$}. Therefore, for $a\leq u \leq b$ we can state that the relation $c \gg u$ holds. That statement enables us to Taylor-expand the term $1/\sqrt{c-u}$,
$$\frac{1}{{\sqrt {c - u} }}   \approx \frac{1}{{\sqrt c }}\left({1 + \frac{u}{{2c}}} \right)$$
and get
$$\frac{{du}}{{\sqrt {\alpha (u - a)(b - u)(c - u)} }} = d\phi\Rightarrow$$
$$ \Rightarrow \frac{1}{{\sqrt {\alpha c} }}\frac{{du}}{{\sqrt{(u - a)(b - u)} }} +\frac{1}{{2c\sqrt {\alpha c}}}\frac{{u\,du}}{{\sqrt {(u - a)(b - u)} }} = d\phi $$
We integrate the last relation for $u$ from $a$ to $b$, so $\phi$ varies from zero to ${{\phi _{{\rm{final}}} }\mathord{\left/ {\vphantom {{\phi _{{\rm{final}}} } {\rm{2}}}}\right. \kern-\nulldelimiterspace} {\rm{2}}}$. The calculations yield
$$\frac{1}{{\sqrt {\alpha c}
}}\int\limits_a^b {\frac{{du}}{{\sqrt {(u - a)(b - u)} }}}  +
\frac{1}{{2c\sqrt {\alpha c} }}\int\limits_a^b
{\frac{{u\,du}}{{\sqrt {(u - a)(b - u)} }}}  =
\int\limits_0^{{{\phi _{{\rm{final}}} } \mathord{\left/
 {\vphantom {{\phi _{{\rm{final}}} } 2}} \right.
 \kern-\nulldelimiterspace} 2}} {d\phi }  \Rightarrow $$
\begin{equation}\label{phi-final}
\Rightarrow \frac{1}{{\sqrt {\alpha c} }}\pi + \frac{1}{{2c\sqrt
{\alpha c} }}\frac{{a + b}}{2}\pi= \frac{{\phi _{{\rm{final}}}
}}{{\rm{2}}}
\end{equation}
for $\int\limits_a^b {\frac{{du}}{{\sqrt {(u - a)(b - u)} }}}  = \left. {\arctan \left( {\frac{{u - \frac{{a + b}}{2}}}{{\sqrt {(u - a)(b - u)} }}} \right)} \right|_{u = a}^{u = b}  = \pi$ and $\int\limits_a^b {\frac{{u\,du}}{{\sqrt {(u - a)(b - u)}}}}=\frac{a+b}{2}\pi $.

Let, now, $a + b = \epsilon  \ll c$. From equation \eqref{coef-p-3} we get $c = \frac{{r_0 }}{{2M}}- \epsilon$ and so
$$\frac{1}{{\sqrt {\alpha c} }}   \approx 1 + \frac{{M\epsilon }}{{r_0 }} + \frac{3}{2}\frac{{M^2 \epsilon^2 }}{{r_0^2 }}$$
Therefore, equation \eqref{phi-final} becomes
$$ \phi _{{\rm{final}}} \approx 2\pi  + 2\pi \frac{{M\epsilon }}{{r_0 }} + 3\pi\frac{{M^2 \epsilon ^2 }}{{r_0^2 }} + \frac{\epsilon }{{2c}}\pi\left( {1 + \frac{{M\epsilon }}{{r_0 }} +\frac{3}{2}\frac{{M^2\epsilon ^2 }}{{r_0^2 }}} \right) $$
where we neglect terms of order higher than $({{M\epsilon }\mathord{\left/ {\vphantom {{M\epsilon } {r_0 }}} \right.
\kern-\nulldelimiterspace} {r_0 }})^2$. Apparently, the resulting precession is
\[
\Delta \phi  = \phi _{{\rm{final}}}  - 2\pi
\]
\[
\Delta \phi  \approx 2\pi \frac{{M\epsilon }}{{r_0 }} + 3\pi
\frac{{M^2 \epsilon ^2 }}{{r_0^2 }} + \frac{\epsilon }{{2c}}\pi
\left( {1 + \frac{{M\epsilon }}{{r_0 }} + \frac{3}{2}\frac{{M^2
\epsilon ^2 }}{{r_0^2 }}} \right)
\]
\[
\Delta \phi  \approx 2\pi \frac{{M\epsilon }}{{r_0 }} + 3\pi
\frac{{M^2 \epsilon ^2 }}{{r_0^2 }} + \frac{\epsilon }{{2\left(
{\frac{{r_0 }}{{2M}} - \epsilon } \right)}}\pi \left( {1 +
\frac{{M\epsilon }}{{r_0 }} + \frac{3}{2}\frac{{M^2 \epsilon ^2
}}{{r_0^2 }}} \right)
\]
\[
\Delta \phi  \approx 2\pi \frac{{M\epsilon }}{{r_0 }} + 3\pi
\frac{{M^2 \epsilon ^2 }}{{r_0^2 }} + \pi \frac{{M\epsilon }}{{r_0
}} \left( {1 + \frac{{2M\epsilon }}{{r_0 }} + \frac{{4M^2 \epsilon
^2 }}{{r_0^2 }}} \right)\left( {1 + \frac{{M\epsilon }}{{r_0 }} +
\frac{3}{2}\frac{{M^2 \epsilon ^2 }}{{r_0^2 }}} \right)
\]
\[
\Delta \phi  \approx 3\pi \frac{{M\epsilon }}{{r_0 }} + 6\pi
\frac{{M^2 \epsilon ^2 }}{{r_0^2 }}
\]
\begin{equation}\label{Delta-phi}
\Delta \phi  \approx 3\pi \frac{{M\epsilon }}{{r_0 }} \left( {1 +
\frac{{2M\epsilon }}{{r_0 }}} \right)
\end{equation}
where we Taylor-expanded the term $\left(\frac{r_0}{2M}-\epsilon\right)^{-1}=\frac{2M}{r_0}\left(1-\frac{2M\epsilon}{r_0}\right)^{-1}$ keeping, one more time, terms of order not higher than $({{M\epsilon } \mathord{\left/ {\vphantom {{M\epsilon } {r_0}}} \right. \kern-\nulldelimiterspace} {r_0 }})^2$.

In order to calculate $\epsilon$ we will use equations \eqref{coef-p-1} and \eqref{coef-p-3}, from which we get
$$\left( {\frac{{r_0 }}{{2M}} - \epsilon } \right)\epsilon  +\frac{{1 - \tilde{E}^2 }}{{2M\tilde{L}^2 }}r_0^3\frac{1}{{\frac{{r_0 }}{{2M}} - \epsilon }} = \frac{{r_0^2
}}{{\tilde{L}^2 }}$$ 
Again, if we use the Taylor expansion of the term $\left(\frac{r_0}{2M}-\epsilon\right)^{-1}$ which appears in the last equation, we take
$$\left[ {1 - \frac{{4M^2 (1 - \tilde{E}^2 )}}{{\tilde{L}^2 }}}\right]\epsilon ^2  - \left( {\frac{{r_0 }}{{2M}} + \frac{{1 -\tilde{E}^2 }}{ {\tilde{L}^2 }}2Mr_0 } \right)\epsilon +\frac{{r_0^2 }} {{\tilde{L}^2 }}\tilde{E}^2  = 0$$ 
The straightforward solution of the last equation gives us the values
\begin{eqnarray}
\epsilon & = & \frac{{\frac{{r_0 }}{{2M}} + \frac{{1 - \tilde{E}^2
}} {{\tilde{L}^2 }}2Mr_0 }}{{2\left[ {1 - \frac{{4M^2 (1 -
\tilde{E}^2 )}}{{\tilde{L}^2 }}} \right]}} \pm {} \nonumber \\
& & {} \frac{{\frac{{r_0 }}{{2M}}\sqrt {1 - \frac{{16M^2
}}{{\tilde{L}^2 }} + \frac{{24M^2 }}{{\tilde{L}^2 }}(1 -
\tilde{E}^2 ) + \frac{{48M^4 }}{{\tilde{L}^4
}}(\tilde{E}^2+\frac{1}{3})(1 - \tilde{E}^2 )} }}{{2\left[ {1 -
\frac{{4M^2 (1 - \tilde{E}^2 )}}{{\tilde{L}^2 }}}
\right]}}\nonumber
\end{eqnarray}
From the two possible values of $\epsilon$ (positive and negative sign) we choose the one with the negative sign. That choice is based on the fact that we need $\epsilon$ to be a small compared to $\frac{r_0}{2M}$ (cf. equation \eqref{coef-p-3} and the discussion that follows it). The negative sign validates the requirement  $\epsilon \ll c\approx\frac{r_0}{2M}$, something that is not accomplished via the use of the positive sign. Therefore, that choice and the fact that for $\tilde{E}^2 \approx 1$ it is $\left({\tilde{E}^2 + \frac{1}{3}} \right)(1- \tilde{E}^2) \approx\frac{4}{3}(1 - \tilde{E}^2)$, enable us to write that
$$\epsilon  \approx \frac{{\frac{{r_0 }}{{2M}} + \frac{{1 -
\tilde{E}^2 }}{{\tilde{L}^2 }}2Mr_0  - \frac{{r_0 }}{{2M}}\sqrt {1
- \frac{{16M^2 }}{{\tilde{L}^2 }} + \frac{{24M^2 }}{{\tilde{L}^2
}}(1 - \tilde{E}^2 ) + \frac{{64M^4 }}{{\tilde{L}^4 }}(1 -
\tilde{E}^2 )} }}{{2\left[ {1 - \frac{{4M^2 (1 - \tilde{E}^2
)}}{{\tilde{L}^2 }}} \right]}}$$
If we Taylor-expand the square root above and keep terms of order not higher than $\frac{M^4}{\tilde{L}^4}$, we get
$$\epsilon  \approx \frac{{r_0 }}{{4M}}\frac{{1 + \frac{{1 -
\tilde{E}^2 }}{{\tilde{L}^2 }}4M^2  - \left[ {1 + \frac{{4M^2
}}{{\tilde{L}^2 }}(1 - 3\tilde{E}^2 ) + \frac{{16M^4
}}{{\tilde{L}^4 }}(3 - 5\tilde{E}^2 )} \right]}}{{ {1 -
\frac{{4M^2 (1 - \tilde{E}^2 )}}{{\tilde{L}^2 }}}}}$$ 
where we neglect the term $\frac{{72M^4 }}{{\tilde{L}^4 }}( {1 -\tilde{E}^2 })^2 $ which appears during the calculations since $\tilde{E}^2 \approx 1$. Finally, we Taylor-expand the term ${\left[ {1 - \frac{{4M^2 (1 - \tilde{E}^2 )}}{{\tilde{L}^2 }}}\right]}^{-1}$
and we end up with
\begin{equation}\label{epsilon}
\epsilon  \approx \left[ {\frac{{2M\tilde{E}^2 }}{{\tilde{L}^2 }}
- \frac{{8M^3 }}{{\tilde{L}^4 }}(\tilde{E}^2  - 3)\left(
{\tilde{E}^2 - \frac{1}{2}} \right)} \right]r_0
\end{equation}
If we replace the acquired value of $\epsilon$ to equation \eqref{Delta-phi} we get the final result
\begin{equation}\label{Dphi(E,L)}
\Delta \phi =\frac{{6\pi M^2 }}{{\tilde{L}^2 }}\left[ {\left( {1 +
\frac{{14M^2 }}{{\tilde{L}^2 }}} \right)\tilde{E}^2 - \frac{{6M^2
}}{{\tilde{L}^2 }}} \right]
\end{equation}
in terms of the total mass of the attractive center and the energy and momentum per unit mass.

Moreover, we can express the precession found (equation \eqref{Dphi(E,L)}) as a function of the eccentricity, $e$, and the semi-major axis, $a$, of the elliptic orbit of the body.  If we assume that the mass of the attractive center is much larger than the mass of the orbiting body, then
$$\tilde{L}^2=Ma(1-e^2) \qquad \textrm{and} \qquad\tilde{E}^2=1-M/2a$$
Therefore the precession in terms of $e$ and $a$ is given by the formula
\begin{equation}\label{Dphi(a,e)}
\Delta \phi  = \frac{{6\pi M}}{{a(1 - e^2 )}}\left[ {1 -
\frac{M}{{2a}}\left( {1 - \frac{{16}}{{1 - e^2 }}} \right)}
\right]
\end{equation}
where we neglect terms of order higher than $(M/a)^2$.

Equation \eqref{Dphi(a,e)} can be used to calculate the precession of the perihelion of planet Mercury's elliptic orbit around the Sun. The semi-major axis of Mercury's orbit is $a=5.791016\times10^{10} \textrm{ m}$, its eccentricity is $e=0.205615$, while the mass of the Sun is $M=1.9892\times10^{30}
\textrm{ kg}$. Of course, in order to find some real numerical values we should write down equation \eqref{Dphi(a,e)} in SI units. Equation \eqref{Dphi(a,e)} as we see it is valid as long as we remember that we use geometrized units, where the speed of light, $c$, and the constant of gravitational attraction, $G$, are taken to be equal to one: $c=G=1$. The form of equation \eqref{Dphi(a,e)} in the SI system of units is
$$\Delta \phi  = \frac{{6\pi GM}}{{ac^2(1 - e^2 )}}\left[ {1 -
\frac{GM}{{2ac^2}}\left( {1 - \frac{{16}}{{1 - e^2 }}} \right)}
\right]$$ 
where $G=6.672599\times10^{-11}\textrm{N}\cdot\textrm{m}^2/\textrm{kg}^2$ and $c=299792458\textrm{ m}/\textrm{s}$. So if we take into account the fact that Mercury spins around the Sun 415 times within a century, then we get the precession
$$\Delta\phi_\textrm{Mercury}=42.964926^{\prime\prime} \textrm{ per century}$$ 
Note that in common bibliography the term $1 - \frac{GM}{2ac^2}\left(1 - \frac{16}{1 - e^2}\right)$---whose numerical value is 1.000004813 in the case of Mercury---is just equal to unity. That exlains the fact that the precession we meet in the bibliography is $\Delta\phi_\textrm{Mercury}=42.964720^{\prime\prime} \textrm{ per century}$. It is clear that the precession derived by this new approach for the case of planet Mercury does not differ much from the precession derived with the usual way \cite{Schutz}. However, if we happen to study the precession of pericenter for an orbit near a strong gravitational field (e.g. in the vicinity of a black hole), then the extra term is expected to produce additional precession that will definitely be significant.

\section{Acknowledgements}
I would like to thank my supervisor at the Nat'l \& Capodistrian University of Athens, dr. T. Apostolatos, for his guidance and support. The above result would have not been found without his help.

\end{document}